\documentclass[11pt]{article}
\usepackage{amsmath,amsfonts, amssymb}
\usepackage{graphicx}   
\usepackage[margin=2.54cm]{geometry}
 
\def\qed{$\Box$}

\newcommand{\R}{\mathbb{R}}

\newcommand{\Rl}{{\mathbb R}}
\newcommand{\A}{{\mathcal A}}
\newcommand{\B}{{\mathcal B}}
\renewcommand{\H}{{\mathcal H}}

\def\idty{{\mathchoice {\mathrm{1\mskip-4mu l}} {\mathrm{1\mskip-4mu l}} %
{\mathrm{1\mskip-4.5mu l}} {\mathrm{1\mskip-5mu l}}}}

\newcommand{\spec}{\mathop{\rm spec}}

\renewcommand{\vec}[1]{\boldsymbol{#1}}

\newcommand{\be}{\begin{equation}}
\newcommand{\ee}{\end{equation}}
\newcommand{\bea}{\begin{eqnarray}}
\newcommand{\eea}{\end{eqnarray}}
\newcommand{\beann}{\begin{eqnarray*}}
\newcommand{\eeann}{\end{eqnarray*}}
\newcommand{\eq}[1]{(\ref{#1})}
\newtheorem{theorem}{Theorem}[section]

\newtheorem{lemma}[theorem]{Lemma}

\begin{document}
\renewcommand{\thefootnote}{\fnsymbol{footnote}}
\title{Locality Estimates for Quantum Spin Systems}
 
\author{Bruno Nachtergaele\\[10pt]
Department of Mathematics\\ University of California at Davis\\
Davis CA 95616, USA\\
\vspace{.3cm} Email: bxn@math.ucdavis.edu\\
\vspace{.3cm} and\\
\vspace{.3cm} Robert Sims \\
Faculty of Mathematics\\ 
University of Vienna\\
1090 Wien, Austria\\
Email:  robert.sims@univie.ac.at}

\date{Version: \today }
\maketitle
\bigskip

\begin{abstract}
We review some recent results that express or rely 
on the locality properties of the dynamics of quantum 
spin systems. In particular, we present a slightly
sharper version of the recently obtained Lieb-Robinson 
bound on the group velocity for such systems on a
large class of metric graphs. Using this bound we
provide expressions of the quasi-locality of the dynamics 
in various forms, present a proof of the Exponential Clustering 
Theorem, and discuss a multi-dimensional Lieb-Schultz-Mattis Theorem.
\end{abstract}

\footnotetext[1]{Copyright \copyright\ 2007 by the authors. 
This paper may be reproduced, in its entirety, for non-commercial
purposes.}

\setcounter{section}{0}

%%%%%%%%%%%%%%%%%%%%%%%%%%%%%%%%%%%%%%%%%%%%%%%%%%%%%%%%%%%
% 
%                      Section 1                          %
%                                                                                        
%%%%%%%%%%%%%%%%%%%%%%%%%%%%%%%%%%%%%%%%%%%%%%%%%%%%%%%%%%%

\section{Introduction}

Locality is a fundamental property of all current physical theories. 
Sets of observables can be associated with points or bounded regions 
in space or space-time and a relativistic dynamics will preserve this 
structure \cite{haag}. How the locality property manifests 
itself mathematically in important situations continues to be an 
active topic of investigation \cite{keyl}.

There is a wide range of important physical systems, however, which 
we prefer to describe by very effective non-relativistic quantum 
theories with Hamiltonian dynamics. Even if the Hamiltonian has 
only finite-range interactions, the dynamics it generates 
generally does not preserve locality, i.e., there is no strict
equivalent to the finite speed of light. However, locality still 
holds in an approximate sense, and there is an associated finite
velocity, which is sometimes referred to as the group 
velocity. We call it the Lieb-Robinson velocity since Lieb and 
Robinson were the first to prove its existence and to obtain a 
bound for it \cite{lieb1972}. They proved that
to a high degree of accuracy locality is preserved 
by quantum spin dynamics in the sense that any local observable 
evolved for a time $t > 0$ remains localized in a region of space 
with diameter proportional to $t$, up to an arbitrarily small
correction. This also means that spatial correlations between 
observables separated by a distance $d$ cannot be established 
faster than a time of order $d$.

The fundamental issue of locality may be sufficient motivation to 
extend the Lieb-Robinson bounds to more general situations, but there 
are other good reasons to try to generalize their result and to
improve the estimates they obtained. As we will discuss below,
locality, or the approximate locality of the dynamics, has been 
shown to be responsible for a considerable number of other important
properties relevant for models of many-body systems. In many
situations, however, the implications of locality have yet to be 
fully explored. 

We will begin this note by presenting a short proof of the new 
Lieb-Robinson bounds obtained successively in \cite{nachtergaele2005},
\cite{hastings2005}, and \cite{NaOgSi}. This improved result we 
give below sharpens the bounds previously obtained in that the
prefactor now only grows as the smallest surface area of the 
supports of the local observables. We do this in Section~\ref{sec:lr}. 
An application where this surface area dependence, rather than 
volume dependence, is important can be found in \cite{eisert2006}.

In Section~\ref{sec:quasilocal}, we present two perspectives on 
how Lieb-Robinson bounds may be used to provide explicit estimates 
on the local structure of the time evolution. As a consequence, 
one easily derives bounds on, for example, multiple
commutators and the rate at which spatial correlations can be 
established in normalized product states. 

Section~\ref{sec:clustering} discusses the so-called 
Exponential Clustering Theorem. In the relativistic context it has 
been known for a long time that a gap in the spectrum above 
the vacuum state implies exponential decay of
spatial correlations in that state 
\cite{araki1962,ruelle1962,fredenhagen85}. That a similar result 
should hold in the non-relativistic 
setting such as quantum spin systems was long expected and taken for 
granted by theoretical physicists \cite{wreszinski87}.  In 
\cite{hastings2004}, Hastings proposed to use Lieb-Robinson 
bounds to obtain such a result  and a complete proof was 
recently given in 
\cite{nachtergaele2005, hastings2005}.

As a final application of these locality bounds, we describe a new proof of the Lieb-Schultz-Mattis
theorem, see \cite{hastings2004, nachtergaele2007}, in Section~\ref{sec:lsm}. 
These results can be traced back to \cite{hastings2004} where Hastings 
introduced a new way to construct and analyze variational 
states for low-lying excitations of gapped Hamiltonians. He developed a notion of a 
{\it quasi-adiabatic evolution} \cite{hastings2005a}
which he then used to 
present a multi-dimensional analogue of the celebrated
Lieb-Schultz-Mattis theorem \cite{liscma}. Such a theorem is 
applicable, for example, to the standard spin-1/2, anti-ferromagnetic 
Heisenberg model and yields an upper bound on the first 
excited state of order $c(\log L)/L$ for systems
of size $L$. His arguments rely on Lieb-Robinson bounds and the Exponential
Clustering Theorem in an essential way, and we have recently obtained a 
rigorous proof of this result which holds in a rather general setting, 
see \cite{nachtergaele2007}. 

We expect that the ideas currently emerging from recent applications of Lieb-Robinson bounds 
will continue to lead to interesting new results for quantum spin systems in the near future.

%%%%%%%%%%%%%%%%%%%%%%%%%%%%%%%%%%%%%%%%%%%%%%%%%%
%
%
%                                 Lieb-Robinson Bounds
%
%
%
%%%%%%%%%%%%%%%%%%%%%%%%%%%%%%%%%%%%%%%%%%%

\section{Lieb-Robinson Bounds} \label{sec:lr}

The first proof of locality bounds in the context of quantum spin
systems appeared in 1972 in a paper by Lieb and Robinson
\cite{lieb1972}.  They proved a bound on the group velocity
corresponding to the dynamics generated by a variety of short range
Hamiltonians. In a series of works  \cite{nachtergaele2005}, \cite{hastings2005}, 
and \cite{NaOgSi}, these estimates have been
generalized, and the proof we provide below, see Theorem~\ref{thm:lr}, 
illustrates many of the new insights which have recently been developed. 

The result stated in Theorem ~\ref{thm:lr} below differs 
from that which may be found in \cite{lieb1972} in two important ways. 
First, the new proof does not require the use of the Fourier 
transform, and therefore, it extends to models defined on sets
without an underlying lattice structure. These results may be of 
interest to those who wish to study quantum spin
systems in the context of quasi-crystals or in the study
of circuits for quantum computation. Second, and most importantly, 
the constants which appear in our bound do not depend on the 
dimensions of the underlying, single-site Hilbert spaces. This
opens up the possibility of applying them to models with
an infinite-dimensional Hilbert space, such as lattice oscillators
\cite{harm}.

The basic set up in this theory concerns quantum spins systems, in
particular, a finite or infinite number of spins labeled by 
$x \in V$. A finite dimensional Hilbert space 
$\mathcal{H}_x$ is assigned to each site $x \in V$. These may
represent the spin of an electron, photon, or an atom. In 
other contexts, these states may represent the ground 
state and first exited state of an atom or a molecule. 
More abstractly, these systems may, for example, model a 
collection qubits, the basic units of quantum information
theory and quantum computation.

If the set $V$ is finite, the Hilbert space of states is given by
$\mathcal{H}_V = \bigotimes_{x \in V} \mathcal{H}_x$. For each spin $x$,
the observables are the complex $n_x \times n_x$ matrices, $M_{n_x}$,
where $n_x = \mbox{dim}( \mathcal{H}_x)$. In this context, the algebra
of observables for the whole system is $\mathcal{A}_V = \bigotimes_{x
  \in V} M_{n_x}$. 

The locality results we wish to describe pertain to observables with
finite support. Here, the support of an observable is understood as
follows. If $X \subset V$, we write $\mathcal{A}_X =
\bigotimes_{x \in X}M_{n_x}$. By identifying $A \in \mathcal{A}_X$
with $A \otimes \idty \in \mathcal{A}_V$, we have that $\mathcal{A}_X
\subset \mathcal{A}_V$. The support of an observable 
$A \in \mathcal{A}_V$ in the minimal set $X \subset V$ for which $A =
A' \otimes \idty$ with $A' \in \mathcal{A}_X$. 

For infinite $V$, the algebra of observables is the completion of
the algebra of local observables given by
$$
\A_V= \bigcup_{X\subset V}\A_X
$$
where the union is over all finite $X \subset V$.

To describe the models we wish to investigate, we must first define
interactions, local Hamiltonians, and the corresponding dynamics.
An interaction is a map $\Phi$ from the set of subsets
of $V$ to $\mathcal{A}_V$ with the property 
that $\Phi(X) \in \mathcal{A}_X$ and $\Phi(X) = \Phi(X)^*$
for all finite $X \subset V$. A quantum spin model 
is defined by a family of local Hamiltonians,
parametrized by finite subsets $\Lambda \subset V$, given by
\begin{equation} \label{eq:defgenham}
H^{\Phi}_{\Lambda} = \sum_{ X \subset \Lambda} \Phi(X).
\end{equation}
For notational convenience, we will often drop the dependence of
$H^{\Phi}_{\Lambda}$ on $\Phi$. The dynamics, or time evolution, 
generated by a quantum spin model is the one-parameter group of 
automorphisms, $\{\tau_t^{\Lambda}\}_{t\in\R}$, defined by
\begin{equation}
\tau_t^{\Lambda}(A)=e^{itH_{\Lambda}} A e^{-itH_{\Lambda}}, \quad A \in \mathcal{A}_{\Lambda},
\end{equation}
which is always well defined for finite sets $\Lambda$. In the context of 
infinite systems, a boundedness condition on the interaction is
required in order for the finite-volume dynamics to converge to 
a strongly continuous one-parameter group of automorphisms on
$\mathcal{A}_V$.

The locality results we prove in Theorem~\ref{thm:lr} are valid for
a large class of interactions. To describe this class precisely,
we first put a condition on the set $V$, which is relevant only in
the event that $V$ is infinite. We assume that the set $V$ is equipped with a
metric $d$ and that there exists a non-increasing function 
$F: [0, \infty) \to (0, \infty)$ for which:

\noindent i) $F$ is uniformly integrable over $V$, i.e.,
\begin{equation} \label{eq:fint}
\| \, F \, \| \, = \, \sup_{x \in V} \sum_{y \in V}
F(d(x,y)) \, < \, \infty,
\end{equation}

\noindent and 

\vspace{.3cm}

\noindent ii) $F$ satisfies
\begin{equation} \label{eq:intlat}
C \, = \, \sup_{x,y \in V} \sum_{z \in V} \frac{F \left( d(x,z) \right) \, F \left( d(z,y)
\right)}{F \left( d(x,y) \right)} \, < \, \infty.
\end{equation} 

Given such a set $V$, it is easy to see that for any $F$ which 
satisfies i) and ii) above, then the family of functions $F_a$, for 
$a \geq 0$, defined by
\begin{equation}
F_a(x) = e^{-ax} \, F(x),
\end{equation}
also satisfy i) and ii) with $\| F_a \| \leq \| F \|$ 
and $C_a \leq C$. In this context, we define the set $\mathcal{B}_a(
V)$ to be those interactions $\Phi$ on $V$ which satisfy
\begin{equation} \label{eq:defnphia}
\| \Phi \|_a \, = \, \sup_{x,y \in V}  \sum_{X \ni x,y} 
\frac{ \| \Phi(X) \|}{F_a \left(
    d(x,y) \right)} \, < \, \infty.
\end{equation}

The Lieb-Robinson bounds we will prove are valid for all $\Phi \in \mathcal{B}_a(V)$
with $a \geq 0$. Simply stated, these results correspond to estimates
of the form
\begin{equation}
\| [ \tau_t^{\Lambda}(A), B] \| \, \leq \, c(A,B) e^{- \mu \left(
    d(A,B) - v_{\Phi} |t| \right)},
\end{equation}
where $A$ and $B$ are local observables, $\tau_t^{\Lambda}(\cdot)$ is the
time evolution corresponding to a finite volume Hamiltonian generated
by an interaction $\Phi \in \mathcal{B}_a(V)$, and $d(A,B)$ is
the distance between the supports of $A$ and $B$. What is crucial in
these estimates is that the constants $c(A,B)$, $\mu$, and $v_{\Phi}$ are
independent of the volume $\Lambda \subset V$ on which
$\tau_t^{\Lambda}( \cdot)$ is defined. Physically, the constant
$v_{\Phi}$ corresponds to a bound on the velocity at which the
dynamics, generated by $\Phi$, propagates through the system.

Intuitively, it is clear that the spread of the interactions through the system should
depend on the surface area of the support of the local observable $A$,
typically denoted by $X$, not it's volume. To make this explicit in 
our bound, we will denote the surface of a set $X$, regarded as a
subset of $\Lambda \subset V$, by 
\begin{equation} \label{eq:defsurf}
S_{\Lambda}(X) \, = \, \left\{ Z \subset \Lambda \, : \, Z \cap X \neq \emptyset
\mbox{ and }  Z \cap X^c \neq \emptyset \right\}.
\end{equation}
Here we will use the notation $S(X) =S_V(X)$, and define 
\begin{equation} \label{eq:npx}
\| \Phi \|_{a}(x;X) \, = \, \left\{ \begin{array}{cc} 
\sup_{y \in \Lambda} \sum_{ \stackrel{Z \in
    S(X):}{x,y \in Z}} \frac{\| \Phi(Z) \|}{F_a \left( d(x,y) \right)}
& \mbox{if } x \in X, \\ 0 & \mbox{otherwise,}
\end{array} \right.
\end{equation}
a quantity corresponding to the interaction terms across the surface of
$X$. Comparing the local quantity appearing in (\ref{eq:npx}) with the norm on the
full interaction given by (\ref{eq:defnphia}), one trivially has that
\begin{equation}
\| \Phi \|_{a}(x; X) \, \leq \, \| \Phi \|_a \, \chi_{\partial_{\Phi}X}(x),
\end{equation}
where we have used $\chi_Y$ to denote the characteristic function of a set 
$Y \subset \Lambda$, and we have introduced the $\Phi$-boundary of a 
set $X$, written $\partial_{\Phi} X$, given by
\begin{equation}
\partial_{\Phi} X \, = \, \left\{ x \in X \, : \, \exists Z \in S(X)
  \mbox{ with } x \in Z \mbox{ and } \Phi(Z) \neq 0 \, \right\}. 
\end{equation}
The Lieb-Robinson bound may be stated as follows.

\begin{theorem}[Lieb-Robinson Bound]\label{thm:lr} 
Let $a \geq 0$ and take $\Phi \in \mathcal{B}_a(V)$. For any finite set 
$\Lambda \subset V$, denote by $\tau_t^{\Lambda}(\cdot)$ the 
time evolution corresponding to the local Hamiltonian 
\begin{equation} \label{eq:lham}
H_{\Lambda} = \sum_{Z \subset \Lambda} \Phi(Z).
\end{equation} 
Given any pair of local observables $A \in
\mathcal{A}_{X}$ and $B \in \A_Y$ with $X, Y \subset
\Lambda$ and $d(X,Y)>0$, one may estimate  
\begin{equation} \label{eq:lrbd1}
\left\| [ \tau_t^{\Lambda}(A), B ] \right\| \, \leq \, \frac{2 \, \| A \|
\, \|B \|}{ \| \Phi \|_a \, C_a} \, \left( e^{2 \, \| \Phi \|_a \,
  C_a \, |t|} - 1 \right) \, D_a(X,Y),
\end{equation}
for any $t \in \mathbb{R}$. Here the function $D_a(X,Y)$ is given by
\begin{equation}
D_a(X,Y) = \min \left[ \sum_{x \in X} \sum_{y \in
  Y} \| \Phi \|_a(x;X) \, F_a \left( d(x,y) \right), \sum_{x \in X} \sum_{y \in
  Y} \| \Phi \|_a(y;Y) \, F_a \left( d(x,y) \right)\right]. 
\end{equation}
\end{theorem}

A number of comments are useful in interpreting this theorem. 
First, we note that if $X$ and $Y$ have a non-empty
intersection, then the argument provided below produces an analogous bound with 
the factor $e^{2 \, \| \Phi \|_a \,C_a \, |t|} - 1 $ replaced by 
$ e^{2 \, \| \Phi \|_a \, C_a \, |t|}  $. In the case of empty intersection 
and for small values of $|t|$, (\ref{eq:lrbd1}) is a better and sometimes
useful estimates than the obvious bound 
$\| [ \tau_t(A), B] \| \leq 2 \| A \| \| B \|$, valid for all 
$t \in \mathbb{R}$.

Next, if $\Phi \in \mathcal{B}_a(V)$ for some $a>0$, then one 
has the trivial estimate that
\begin{equation} \label{eq:dbd}
D_a(X,Y) \, \leq \, \| F_0 \| \, \| \Phi \|_a \, \min \left( \left|
    \partial_{\Phi}X \right|, \left| \partial_{\Phi}Y \right| \right) \, e^{-a \, d(X,Y)}.
\end{equation}
Clearly then, we have that   
\begin{equation} \label{eq:vel}
\left\| [ \tau_t^{\Lambda}(A), B ] \right\| \, \leq \, \frac{2 \, \| A \|
\, \|B \| \, \|F_0 \|}{ C_a} \, \min \left( \left|
    \partial_{\Phi}X \right|, \left| \partial_{\Phi}Y \right| \right) \, e^{- a \,\left[
 d(X,Y) - \frac{2 \| \Phi \|_a C_a}{a} |t| \right]},
\end{equation}
which corresponds to a bound on the velocity of propagation given by
\begin{equation}
v_{\Phi} \leq \inf_{a>0} \frac{2 \| \Phi \|_a C_a}{a}.
\end{equation}

Next, we observe that  for fixed local observables $A$ and $B$, the 
bounds above, (\ref{eq:lrbd1}) and (\ref{eq:vel}), are 
independent of the volume $\Lambda \subset V$; given that $\Lambda$
contains the supports of both $A$ and $B$. Furthermore, we note that  
these bounds only require that one of the observables has 
finite support; in particular, if  $|X|< \infty$ and $d(X,Y)>0$, 
then the bounds are valid irrespective of the support of $B$.  

%%%%%%%%%%%%%%%%%%%%%%%%%%%%%%%%%%%%%%%%%%%
%%%%%%%%%%%%%%%%%%%%%%%%%%%%%%%%%%%%%%%%%%
{\it Proof of Theorem~\ref{thm:lr}:}
To prove (\ref{eq:lrbd1}), we will provide an estimate on the quantity
\begin{equation} \label{def:cb}
C_B(Z; t)   = \sup_{A \in \mathcal{A}_Z} \frac{
  \| [ \tau_t^{\Lambda}(A), B] \|}{ \| A \|},
\end{equation} 
where $B$ is a fixed observable with support in $Y$, and the subset 
$Z \subset \Lambda$ we regard as arbitrary. Introduce the function
\begin{equation}
f(t) = \left[ \tau_t^{\Lambda} \left( \tau_{-t}^X(A) \right), B \right],
\end{equation}
where $A$ and $B$ are as in the statement of the theorem.
Due to the strict locality of the Hamiltonian $H_X$, as defined e.g. in
(\ref{eq:lham}) and the fact that the observable $A \in
\mathcal{A}_X$, we have that $\mbox{supp}(\tau_{-t}^X(A)) \subset X$ for all
$t \in \mathbb{R}$. It is straight forward to verify that
\begin{equation} \label{eq:derf}
f'(t) = i \sum_{Z \in S_{\Lambda}(X)} \left[ \tau_t^{\Lambda} \left( \Phi(Z)
    \right), f(t) \right] - i  \sum_{Z \in S_{\Lambda}(X)} \left[
    \tau_t^{\Lambda}( \tau_{-t}^X(A)), \left[ \tau_t^{\Lambda} \left(
      \Phi(Z) \right), B \right] \right] .
\end{equation}
Since the first term in (\ref{eq:derf}) is norm preserving, we find
that
\begin{equation} \label{eq:nlrb1}
\| [ \tau_t^{\Lambda}( \tau_{-t}^X(A)), B] \| \, \leq  \, \| [A,B] \|
\,  + \,  2 \, \| A \| \, \sum_{Z \in S(X)} \int_0^t 
    \left\| [ \tau_s^{\Lambda}( \Phi(Z)), B] \right\| ds.
\end{equation}
The inequality  (\ref{eq:nlrb1}) and the fact that $\|
\tau_{-t}^X(A) \| = \| A \|$ together imply that 
\begin{equation} \label{eq:supest}
C_B(X;t) \, \leq \, C_B(X;0) \, + \, 2 \sum_{Z \in S(X)} 
\| \Phi(Z) \| \int_0^{|t|} C_B(Z; s) ds.
\end{equation}
It is clear from the definition, see (\ref{def:cb}), that for any
finite $Z \subset \Lambda$,
\begin{equation}
C_B(Z; 0) \, \leq \, 2 \, \| B \| \, \delta_Y(Z)
\end{equation}
where $\delta_Y(Z) = 0$ if $Z \cap Y = \emptyset$ and $\delta_Y(Z) =
1$ otherwise.  Using this fact, iteration of (\ref{eq:supest}) yields that 
\begin{equation}  \label{eq:seriesbd}
C_B(X,t) \, \leq \, 2 \| B \| \, \sum_{n=0}^{ \infty}
\frac{(2|t|)^n}{n!} a_n,
\end{equation}
where for $n \geq 1$,
\begin{equation}
a_n \, = \, \sum_{Z_1 \in S(X)} \sum_{Z_2 \in S(Z_1)} \cdots \sum_{Z_n
  \in S(Z_{n-1})} \delta_Y(Z_n) \, \prod_{i=1}^n \| \Phi(Z_i) \| \, .
\end{equation}

For an interaction $\Phi \in \mathcal{B}_a(V)$, one may estimate that
\begin{equation} 
a_1 \, \leq \, \sum_{x \in X} \sum_{y \in Y} \sum_{\stackrel{Z \in
    S(X):}{x, y \in Z}} \| \Phi(Z) \| \, \leq \, \sum_{x \in X}
\sum_{y \in Y} \| \Phi \|_{a}(x;X) F_a \left( d(x, y) \right).
\end{equation}
In addition,
\begin{eqnarray} 
a_2 & \leq & \sum_{x \in X} \sum_{y \in Y} \sum_{z \in \Lambda}
\sum_{ \stackrel{Z_1 \in S(X):}{x, z \in Z_1}} 
\| \Phi(Z_1) \| \sum_{ \stackrel{Z_2 \in S(Z_1):}{z, y \in Z_2}}  \| \Phi(Z_2) \|  \nonumber \\ 
& \leq &   \| \Phi \|_a \, \sum_{x \in X} \sum_{y \in Y} \sum_{z \in
  \Lambda} F_a \left( d(z, y) \right)  \, \sum_{ \stackrel{Z_1 \in S(X):}{x, z \in Z_1}} 
\| \Phi(Z_1) \|  \nonumber \\ & \leq &   \| \Phi \|_a \, \sum_{x \in X} \sum_{y \in Y} \sum_{z \in
  \Lambda} F_a \left( d(x,z) \right) \, F_a \left( d(z, y) \right)  \,
\| \Phi \|_a(x;X) \nonumber \\
 & \leq & \| \Phi \|_a \, C_a \, \sum_{x \in X} \sum_{y \in
  Y} \| \Phi \|_a(x; X) \, F_a \left( d(x,y) \right),
\end{eqnarray}
where we have used (\ref{eq:intlat}) for the final inequality. 
With analogous arguments, one finds that for all $n \geq 1$,
\begin{equation} \label{eq:aneq}
a_n \, \leq \,  \left( \| \Phi \|_a \, C_a \right)^{n-1} \, \sum_{x \in X} \sum_{y \in
  Y} \| \Phi \|_a(x; X) \, F_a \left( d(x,y) \right).
\end{equation}
Inserting (\ref{eq:aneq}) into (\ref{eq:seriesbd}) we see that
\begin{equation} \label{eq:lrbdd}
C_B(X,t) \, \leq \,  \frac{2 \, \| B \| }{\|
  \Phi \|_a \, C_a} \, \left( e^{2 \, \| \Phi \|_a \, C_a \, |t|} - 1
\right) \, \sum_{x \in X} \sum_{y \in Y} \| \Phi \|_a(x;X) \, F_a \left( d(x,y) \right),
\end{equation}
from which (\ref{eq:lrbd1}) immediately follows. 
\hfill \qed

%%%%%%%%%%%%%%%%%%%%%%%%%%%%
%
%
%
%     Quasi-Locality
%
%
%
%%%%%%%%%%%%%%%%%%%%%%%%%%%%%%

\section{Quasi-locality  of the dynamics}\label{sec:quasilocal}

The Lieb-Robinson bounds of Theorem~\ref{thm:lr} imply that the dynamics
of quantum lattice systems, those generated by short range 
interactions, are {\it quasi-local} in the sense that the diameter of 
the support of any evolved local observable does not grow faster than 
linearly with time, up to an arbitrarily small error. There are at 
least two interesting ways to give precise meaning to this quasi-
locality property of the dynamics. In the first, one shows that
the time-evolved observable can be well-approximated in norm by one 
with a strictly local support. This is achieved by the quantum version 
of integrating out the variables in the complement of a ball with a 
radius proportional to time. In the second, we show that to compute 
the dynamics up to a time $t>0$, one can replace the Hamiltonian with a 
local Hamiltonian supported in a ball of radius proportional to $t$. 
Clearly, the net result is the same: the support of observables evolved 
with approximate dynamics remains contained in the ball where
the local Hamiltonian is supported. After presenting the details of 
these two approaches, we conclude this section with a few interesting 
applications that immediately follow from quasi-locality.

As in the previous section, we will work with an interaction 
$\Phi \in \mathcal{B}_a(V)$ with $a>0$. For the purpose of the 
discussion below, we will consider a 
finite subset $\Lambda \subset V$ and restrict our attention to the
dynamics $\tau^{\Lambda}_t( \cdot)$ generated by the finite volume
Hamiltonian $H_{\Lambda}$ as defined in (\ref{eq:defgenham}); our bounds will be independent of the
volume $\Lambda$. For any $X \subset \Lambda$ we will denote 
by $X^c = \Lambda \setminus X$. 

In the first approach to obtaining a local approximation with support contained in $X\subset\Lambda$, one takes the normalized partial trace
over the Hilbert space associated with $X^c$. In order to estimate
the norm difference it is convenient to calculate the partial
trace as an integral over the group of unitaries
\cite{bravyi2006}. Given an arbitrary observable 
$A \in \A_{\Lambda}$ and a set $X \subset \Lambda$, define 
\begin{equation}
\langle A \rangle_{X^c} = \int_{\mathcal{U}(X^c)} U^* A U \, \mu(dU),
\end{equation}
where $\mathcal{U}(X^c)$ denotes the group of unitary operators over
the Hilbert space $\mathcal{H}_{X^c}$ and $\mu$ is the associated, 
normalized Haar measure. It is easy to see that for any 
$A \in \A_{\Lambda}$, the quantity $\langle A
\rangle_{X^c}$ has been localized to $X$ in the sense that $\langle A
\rangle_{X^c} \in \A_X$. Moreover, the difference may be written in
terms of a commutator, i.e.  as
\begin{equation} \label{eq:acomm}
\langle A \rangle_{X^c} \, - \, A \, = \, 
\int_{\mathcal{U}(X^c)} U^* \left[ A,  U  \right] \, \mu(dU).
\end{equation}

To localize the dynamics, let $A \in \mathcal{A}_X$
and fix $\epsilon >0$. Based on our estimates in Theorem~\ref{thm:lr},
we approximate the support of $\tau^{\Lambda}_t(A)$ with a
time-dependent ball 
\begin{equation} \label{eq:defball}
B_t( \epsilon, A) \, = \,  \left\{ x \in \Lambda \, : \, d(x, X) \, \leq \,
  \frac{2 \| \Phi \|_a C_a}{a} \, |t| \, + \, \epsilon \, \right\}.
\end{equation} 
For any unitary $U \in \mathcal{U}(B_t(\epsilon, A)^c)$, we clearly have that
\begin{equation}
d \left( X, {\rm supp}(U) \right) \, \geq \, \frac{2 \| \Phi \|_a
  C_a}{a} \,|t| \, + \epsilon,
\end{equation}
and therefore, using (\ref{eq:acomm}) above and our bound (\ref{eq:vel}), we immediately
conclude that 
\begin{eqnarray} \label{eq:onetru}
\left\| \, \tau_t^{\Lambda}(A) \, - \, \langle \tau_t^{\Lambda}(A) \rangle_{B_t(\epsilon,A)^c} \, \right\| & 
\leq & \int_{\mathcal{U}(B_t^c(\epsilon))}  \left\| \,  \left[ \tau_t^{\Lambda}(A),  U
  \right] \, \right\| \, \mu(dU) \nonumber \\
& \leq & \frac{2 \, \| A \| \, \left| \partial_{\Phi} X \right| }{C_a} \, \| F_0 \| \,
e^{- a \epsilon}.
\end{eqnarray}
The tolerance $\epsilon>0$ can be chosen to optimize estimates.

In the second approach, one shows that only those terms in the 
Hamiltonian supported in $B_T(\epsilon,A)$ contribute significantly to
the time evolution of $A$ up to time $T$ \cite{NaOgSi}. Again, we consider
the finite volume dynamics applied to a local
observable $A \in \mathcal{A}_X$. Fix $\epsilon >0$, $T>0$, and
consider the ball $B_T( \epsilon, A)$ as defined in (\ref{eq:defball})
above. The estimate
\begin{equation} \label{eq:trunc2}
\left\| \tau_t^{\Lambda}(A) \, - \, \tau_t^{B_T( \epsilon, A) }(A) \right\| \, \leq
\, \frac{ \| A \| \, \| F_0 \| \, | \partial_{\Phi} X |}{ C_a^2} \, \left( C_a + \| F_a \| \right) \, e^{-a \epsilon},
\end{equation} 
valid for all $|t| \leq T$, readily follows from the results in \cite{NaOgSi}.
In fact, the proof of (\ref{eq:trunc2}) uses the following basic estimate, see
e.g. Lemma 3.3 in \cite{NaOgSi}, 
\begin{lemma} \label{lem:intdrop} Let $\Phi \in \mathcal{B}_a(V)$ with
  $a>0$ and take a finite subset $\Lambda \subset V$. If the finite
  volume Hamiltonian is written as the sum of two self-adjoint
  operators, $H_{\Lambda} = H_{\Lambda}^{(1)} + H_{\Lambda}^{(2)}$,
  and for $i=1,2$, $\tau_t^{(i)}( \cdot)$ denotes the dynamics
  corresponding to $H_{\Lambda}^{(i)}$, then the following estimate is
  valid:
\begin{equation}
\left\| \tau_t^{\Lambda}(A) \, - \, \tau_t^{(1)}(A) \right\| \, \leq
\, \int_0^{|t|} \left\| \left[ H_{\Lambda}^{(2)}, \tau_s^{(1)}(A) \right] \right\| \, ds,
\end{equation}
for any observable $A$ and $t \in \mathbb{R}$.
\end{lemma}

To apply Lemma~\ref{lem:intdrop} in the context discussed above, we 
write the local Hamiltonian as the sum of two terms:
\begin{equation} \label{eq:hamdecomp}
H_{\Lambda} \, = \, \sum_{\stackrel{Z \subset \Lambda:}{Z \notin
    S_{\Lambda} \left( B_T( \epsilon, A) \right)}} \Phi(Z) \, + \, 
\sum_{\stackrel{Z \subset \Lambda:}{Z \in
    S_{\Lambda} \left( B_T( \epsilon, A) \right)}} \Phi(Z) \, = \,
H_{\Lambda}^{(1)} \, + \, H_{\Lambda}^{(2)}. 
\end{equation}
Recall that for any $X \subset \Lambda$, we defined $S_{\Lambda}(X)$,
the surface across $X$ in $\Lambda$, with equation (\ref{eq:defsurf}). 
Dropping the surface terms comprising $H_{\Lambda}^{(2)}$ 
above, decouples the dynamics, i.e.,  
$\tau_t^{(1)}(A) = \tau_t^{B_T( \epsilon,A)}(A)$, and we find that for
any $|t| \leq T$,
\begin{equation}
\left\| \tau_t^{\Lambda}(A) \, - \, \tau_t^{B_T( \epsilon, A) }(A) \right\| \, \leq
\, \sum_{\stackrel{Z \in \Lambda:}{ Z \in S_{\Lambda} \left(B_T( \epsilon, A) \right)}}
  \int_0^{|t|} \left\| \left[ \Phi(Z), \tau_s^{B_T( \epsilon, A)}(A) \right] \right\| \, ds.
\end{equation}
For each of the terms on the right hand side above, the Lieb-Robinson
estimates imply that
\begin{equation}
 \left\| \left[ \Phi(Z), \tau_s^{B_T( \epsilon, A)}(A) \right]
 \right\| \, \leq \, \frac{2 \, \| \Phi(Z) \| \, \| A \|}{C_a} e^{ 2 \,
   \| \Phi \|_a \, C_a \, |s|} \, \sum_{z \in Z} \sum_{x \in
   \partial_{\Phi}X} F_a \left( d(x,z) \right).
\end{equation}
Careful consideration of the combinatorics of these sums, exactly as
done in \cite{NaOgSi}, yields (\ref{eq:trunc2}) as claimed. 

It is interesting to ask about locality bounds for complex times.
For one-dimensional systems with finite range interactions, Araki proved 
that the support does not grow faster than an exponential in $\vert z\vert$,
where $z$ is the complex time \cite[Theorem 4.2]{araki1969}. In other words,
the complex time evolution
$$
\tau^{[-R,R]}_z(A)=e^{i z H_\Lambda}Ae^{-i z H_\Lambda}
$$
can be approximated by $\tau^{[-N,N]}_z(A)$ with small error, uniformly 
in $R$, as long as $\log \vert z\vert\leq cN$, for a suitable constant $c$.
For stochastic dynamics of classical particle systems, good locality
bounds that are very similar to Theorem 
\ref{thm:lr} are known \cite[Theorem 4.20]{liggett1985}. 
Since these classical models with stochastic dynamics can be thought
of as quantum systems, generated by particular Hamiltonians and 
evolving with purely imaginary times, this may indicate that the general result,
which allows for no more than logarithmic growth of $\vert z\vert$, 
should not be considered the final word on this issue in regards to 
specific models of interest. 

The quasi-locality formulation of the Lieb-Robinson bounds makes it
easy to derive bounds on double and higher order commutators of the
form
$$
[\tau_{t_1} (A) ,[\tau_{t_2}(B),\tau_{t_3}(C]]\, .
$$
Such commutators remain small in norm as long as $t_1,t_2$, and $t_3$ are
such that the intersection of the time-dependent ``quasi-supports'' 
remains empty. That is whenever $t_1, t_2, t_3\in\Rl$ are such that 
$$
B_{t_1}(\epsilon, A)\cap B_{t_2}(\epsilon, B)\cap B_{t_3}(\epsilon, C)=\emptyset\, .
$$

Another immediate application of the quasi-locality property 
is a bound on the rate at which spatial correlations can be
created by the dynamics starting from a product state. 
In \cite{NaOgSi}, we proved the following.
\begin{theorem}
Let $a>0$, $\Phi \in \mathcal{B}_a(V)$, and take $\Omega$ to be a
normalized product state. Given $X, Y \subset \Lambda$ with $d(X,Y)>0$
and local observables $A \in \mathcal{A}_X$ and $B \in \mathcal{A}_Y$,
the estimate
\begin{equation}
\left| \langle \tau_t(AB) \rangle \, - \, \langle \tau_t(A) \rangle
  \, \langle \tau_t(B) \rangle \right| \, \leq \, 4 \, \| A \| \, \|
B \| \, \| F \| \left( \left| \partial_{\Phi} X \right| \, + \, \left|
    \partial_{\Phi}Y \right| \right) \, G_a(t) \, e^{- a d(X,Y)} 
\end{equation}
is valid for all $t \in \mathbb{R}$. Here, for any observable $A$, the
expectation value of $A$ in the state $\Omega$ is denoted by 
$\langle A \rangle = \langle \Omega, A \Omega \rangle$, and the
function
\begin{equation}
G_a(t) \, = \, \frac{C_a + \| F_a \|}{C_a} \, \| \Phi \|_a \,
\int_0^{|t|} e^{2 \| \Phi \|_a C_a |s|} \, ds.
\end{equation}
\end{theorem}

%%%%%%%%%%%%%%%%%%%%%%%%%%%%%%%%
%
%
%
%                                    Exponential Clustering
%
%
%%%%%%%%%%%%%%%%%%%%%%%%%%%%%%%%%%%

\section{Exponential Clustering} \label{sec:clustering}

As a second application of these locality bounds, we will present a
proof of the Exponential Clustering Theorem, see
Theorem~\ref{thm:decay} below, which improves on the estimates found
in \cite{nachtergaele2005}, see also \cite{hastings2005}. The proof of 
exponential clustering demonstrates that models with a spectral gap 
above the ground state energy necessarily
exhibit exponential decay of spatial correlations in their ground state. Such
results have recently appeared in a variety of contexts, e.g. \cite{nachtergaele2005, hastings2005}.
Using valence bond states, as is done in \cite{nachtergaele1996}, one can easily
construct gapless models with exponentially decaying ground state
correlations indicating that, in general, there is no converse to
Theorem~\ref{thm:decay}. 

In the finite volume, the notion of a gapped Hamiltonian is clear.
If the system is infinite, we express the gap condition in terms of the
limiting dynamics, the existence of which is guaranteed by the Lieb-Robinson bounds
as discussed above, by considering a representation of the system on a Hilbert space 
$\mathcal{H}$.  This means that there
is a representation $\pi:\A_V \to \B(\H)$, and a self-adjoint operator $H$ on
$\H$ such that
\begin{equation} \label{eq:rep}
\pi(\tau_t(A))=e^{itH}\pi(A)e^{-it H},\quad A\in \A_V .
\end{equation}
We will assume that the representative 
operator $H$ is non-negative and that there exists a vector $\Omega\in\H$
for which $H\Omega =0$. We say that the system has a spectral
gap in the representation if there exists $\delta >0$ such that $\spec (H) \cap
(0,\delta) =\emptyset$. In this case, the spectral gap, $\gamma$, is defined
by
\begin{equation} \label{eq:gap}
\gamma=\sup\{\delta > 0 \mid \spec(H) \cap (0,\delta) =\emptyset\}.
\end{equation}
Let $P_0$ denote the orthogonal projection onto $\ker H$. From now on, we will work in this
representation and simply write $A$ instead of
$\pi(A)$.

%We will derive a bound for ground state correlations of the form
%\be
%\langle\Omega, A\tau_{ib}(B)\Omega\rangle
%\ee
%where $b \geq 0$ and $A$ and $B$ are local observables. 
%The case $b=0$ is the standard (equal-time) correlation
%function.

\begin{theorem}[Exponential Clustering]\label{thm:decay}
Let $a>0$ and take $\Phi \in \mathcal{B}_a(V)$. Suppose that the
dynamics corresponding to $\Phi$ on $V$ can be represented by a
Hamiltonian $H$ with a gap $\gamma >0$ above the ground state energy,
as described above. Let $\Omega$ be a normalized ground state vector
for $H$; i.e. satisfy $H \Omega = 0$ with $\| \Omega \| = 1$. Then,
there exists a constant $\mu > 0$ such that for any
local observables $A\in\A_X$ and $B\in\A_Y$ with $X, Y \subset V$ and $d(X,Y) >0$ 
satisfying $P_0 B\Omega = P_0 B^*\Omega =0$, the bound 
\begin{equation} \label{decay}
\left\vert \langle\Omega, A \tau_{ib}(B) \Omega \rangle \right\vert \, \leq \, C(A,B, \gamma) \, e^{- \mu d(X,Y) \left( 1 + \frac{ \gamma^2 b^2}{4 \mu^2 d(X,Y)^2} \right)}
\end{equation}
%\begin{equation} \label{decay}
%\frac{\left\vert \langle\Omega, A\tau_{ib}(B) \Omega\rangle
 % \right\vert}{ \| A \| \, \|B \|} \, \leq \, \left[ 1 \, + \, \sqrt{
  %  \frac{1}{ \mu d(X,Y)}} \, + \, \frac{2 \| F_0 \|}{ \pi \| \Phi
  %  \|_a C_a} \, \min \left( S_a(X), S_a(Y) \right) \, \right] \, 
%e^{-\mu d(X,Y) g(b)} 
%\end{equation}
is valid for all non-negative $b$ satisfying $0 \leq b \gamma \leq 2 \mu d(X,Y)$. 
One may take
\begin{equation}
\mu \, = \, \frac{a \, \gamma }{4 \Vert \Phi \Vert_a C_a  \, + \,
  \gamma},
\end{equation}
as well as a constant
\begin{equation}
C(A,B, \gamma) \, = \, \| A \| \, \| B \| \, \left[ 1 \, + \, \sqrt{
    \frac{1}{ \mu d(X,Y)}} \, + \, \frac{2 \| F_0 \|}{ \pi C_a} \, \min \left( | \partial_{\Phi} X|, | \partial_{\Phi} Y| \right) \, \right].
\end{equation} 

\end{theorem}

Note that in the case of a non-degenerate ground state, the condition on $B$ is equivalent
to $\langle\Omega, B\Omega\rangle=0$. In this case, the theorem with
$b=0$ becomes
\be \label{zerob}
\left\vert \langle\Omega, AB\Omega\rangle
-  \langle\Omega, A\Omega\rangle\, \langle\Omega, B\Omega\rangle\right\vert 
\leq C(A,B, \gamma) \, e^{-\mu d(X,Y)} ,
\ee
which is the standard (equal-time) correlation function. For small $b>0$, the 
estimate \eq{decay} can be viewed as a perturbation of \eq{zerob}. Moreover, 
for $b>0$ large, there is a trivial bound
\be
\left| \langle \Omega, A\tau_{ib}(B) \Omega \rangle \right| \, \leq \, \| A\| \, \| B \| \, e^{-b\gamma} .
\ee

%%%%%%%%%%%%%%%%%%%%%%%%%%%%%%%%%%%%%%%%%%%%%%%%%%%%%%%%%%

{\it Proof of Theorem~\ref{thm:decay}:}

We will follow very closely the proof which appears in \cite{nachtergaele2005} and
refer to it whenever convenient. Consider the function $f$ given by
\begin{equation} \label{eq:deff}
f(z) \, = \, \left\langle \Omega, A \tau_z(B) \Omega \right\rangle \,
= \, \int_{\gamma}^{\infty} e^{izE} \, d \left\langle A^* \Omega, P_E
  B \Omega \right\rangle,
\end{equation}
where we have used the spectral theorem and the fact that $B$ projects
off the ground state. It is clear that the function $f$ defined in
(\ref{eq:deff}) is analytic in the upper half plane and has a
continuous (and bounded) boundary value on the real axis. 
The quantity we wish to bound corresponds to $f(ib)$
for $b>0$. The case $b=0$ will follow by a limiting argument.

Since the boundary value of $f$ on $\mathbb{R}$ is continuous, one may
show by a limiting argument that for any $T>b$,
\begin{equation}
f(ib) \, = \, \frac{1}{2 \pi i} \int_{\Gamma_T} \frac{f(z)}{z-ib} dz,
\end{equation}
where $\Gamma_T$ is the semi-circular contour from $-T$ to $T$ (on the
real axis) into the upper half plane. As is shown in \cite{nachtergaele2005}, the fact
that the Hamiltonian is gapped, see also (\ref{eq:deff}), implies that
the integral over the circular part of the contour vanishes in the
limit as $T \to \infty$, and therefore we have the bound
\begin{equation} \label{eq:fintbd}
 \left| \left\langle \Omega, A \tau_{ib}(B) \Omega \right\rangle
 \right|  \, = \, |f(ib)| \, \leq \, \limsup_{T \to \infty} \left|
   \frac{1}{2 \pi i} \int_{-T}^T \frac{f(t)}{t-ib} dt \, \right|. 
\end{equation}
The proof is then complete once we estimate this integral over the
real line. While this inequality is true for any value of $b>0$, to 
get the desired estimate, we will have to choose $b>0$ sufficiently 
small; see the comments following (\ref{eq:defs}) below.

Let $\alpha >0$. We will choose this free parameter later.
Observe that one may write
\begin{equation}
f(t) \, = \, e^{- \alpha b^2} \, \left[ f(t) e^{ - \alpha t^2} \,
  + \, f(t) \left( e^{ \alpha b^2} - e^{ - \alpha t^2} \right) \right].
\end{equation}
Clearly then, the integral we wish to bound, the one on the 
right hand side of (\ref{eq:fintbd}) above, can be estimated by
\begin{equation} \label{eq:intbd1}
 e^{ - \alpha b^2} \, \left| \frac{1}{2
   \pi i} \int_{-T}^T \frac{f(t) e^{- \alpha t^2}}{t - ib} dt
\right| \, + \,  e^{ - \alpha b^2} \, \left| \frac{1}{2
   \pi i} \int_{-T}^T \frac{f(t) \left( e^{ \alpha b^2} - e^{- \alpha t^2}\right)}{t - ib} dt \right|.
\end{equation}
We will bound the absolute value of each of the integrals appearing
in (\ref{eq:intbd1}) separately; the prefactor $e^{- \alpha b^2}$ will be an
additional damping made explicit by the choice of $\alpha$.

To bound the first integral appearing in (\ref{eq:intbd1}), 
we further divide the integrand into two terms. Note that
\begin{equation} \label{eq:f0decomp}
f(t) \, e^{- \alpha t^2} \, = \, \langle \Omega, \tau_t(B)A \Omega
\rangle e^{- \alpha t^2} \, +
 \, \langle \Omega, \left[ A, \tau_t(B) \right] \Omega \rangle e^{- \alpha t^2}.
\end{equation}
By the spectral theorem, we have that
\begin{equation}
\frac{1}{2 \pi i} \, \int_{-T}^T \frac{  \langle \Omega, \tau_t(B)A \Omega
\rangle \, e^{- \alpha t^2}}{t - ib} dt \, = \, \int_{\gamma}^{\infty}
\frac{1}{2 \pi i} \int_{-T}^T \frac{e^{-itE} e^{- \alpha t^2}}{t- ib}
dt \, d \langle P_E B^* \Omega, A \Omega \rangle, 
\end{equation}
where we have used now that $B^* \Omega$ is also orthogonal to the ground
state. Applying Lemma~\ref{lem:ft}, stated below, to the inner integral above, we
have that
\begin{equation}
\lim_{T \to \infty} \frac{1}{2 \pi i} 
\int_{-T}^T \frac{e^{-itE} \, e^{- \alpha t^2}}{t - ib} dt \,
= \, \frac{1}{2 \sqrt{ \pi \alpha}} \int_0^{\infty} e^{-bw} e^{-
  \frac{(w+E)^2}{4 \alpha}} dw \, \leq \, \frac{1}{2} e^{ -
  \frac{\gamma^2}{4 \alpha}},
\end{equation}
where for the final inequality above we used that $E \geq \gamma >0$,
$\alpha >0$, and $b>0$. From this we easily conclude that
\begin{equation}  \label{eq:bd1}
\limsup_{T \to \infty}  \left| \frac{1}{2
   \pi i} \int_{-T}^T \frac{\langle \Omega, \tau_t(B)A \Omega
\rangle \, e^{- \alpha t^2}}{t - ib} dt \right| \, \leq \, \frac{ \| A
\| \, \| B \|}{2} e^{ - \frac{\gamma^2}{4 \alpha}}.
\end{equation}

For the integral corresponding to the second term in
(\ref{eq:f0decomp}), it is easy
to see that
\begin{equation} \label{eq:intbd2}
 \left| \frac{1}{2 \pi i} \int_{-T}^T \frac{\langle \Omega, \left[ A,
       \tau_t(B) \right] \Omega \rangle \, e^{- \alpha t^2}}{t - ib} 
dt \right| \, \leq \, \frac{1}{2 \pi} \int_{- \infty}^{\infty}
\frac{ \| [ A, \tau_t(B) ] \|}{|t|} e^{- \alpha t^2} dt,
\end{equation}
where we have taken advantage of the fact that $b>0$. To complete our 
estimate, we will introduce another free parameter $s>0$. Here we use 
the Lieb-Robinson bound, Theorem~\ref{thm:lr}, for times $|t| \leq s$ and
a basic norm estimate otherwise. The result is that the right hand
side of (\ref{eq:intbd2}) is bounded from above by
\begin{equation} \label{eq:bd2}
\frac{2 \, \| A \| \, \| B \|}{ \pi \, \| \Phi \|_a \, C_a }
D_a(X,Y) \left( e^{2 \| \Phi \|_a C_a s} - 1 \right) \, + \, \frac{ \|
  A \| \, \| B \|}{s \sqrt{ \pi \alpha}} e^{- \alpha s^2}. 
\end{equation}
This completes the bound of the first integral appearing in (\ref{eq:intbd1}).

Using again the spectral theorem, the second integral in (\ref{eq:intbd1}) may
be written as
\begin{equation} \label{eq:moreint}
 \left|  \int_{\gamma}^{\infty}
\frac{1}{2 \pi i} \int_{-T}^T \frac{e^{itE} \left(e^{ \alpha b^2} -
    e^{- \alpha t^2} \right)}{t- ib}
dt \, d \langle  A^* \Omega, P_E B \Omega \rangle \right|.
\end{equation}
As is described in detail in \cite{nachtergaele2005}, we find
that for $E \geq \gamma$ and $\alpha$ chosen such that $\gamma \geq 2 \alpha b$,
\begin{equation} \label{eq:bd3}
\lim_{T \to \infty} \frac{1}{2 \pi i} \int_{-T}^T \frac{e^{itE}}{t-
  ib} \left( e^{\alpha b^2} - e^{- \alpha t^2} \right) dt \, \leq \,
\frac{1}{2} e^{ - \frac{ \gamma^2}{4 \alpha}},
\end{equation} 
which produces an estimate (one analogous to the bound in
(\ref{eq:bd1}) above) for (\ref{eq:moreint}).

All of our estimates above combine to demonstrate that the right 
hand side of (\ref{eq:fintbd}) is bounded by  
\begin{equation}
\| A \| \, \| B \| \, \left[ e^{ - \frac{ \gamma^2}{4 \alpha}} \, + \,
 \frac{2 \, D_a(X,Y)}{ \pi \, \| \Phi \|_a \, C_a} \left( 
e^{2 \, \| \Phi \|_a \, C_a \, s} - 1 \right) \, + \, \frac{1}{s \sqrt{\pi\alpha}} e^{- \alpha s^2} \right]
\end{equation}
 if $\alpha$ satisfies $\gamma \geq 2 \alpha b$.
The choice $\alpha = \gamma / 2s$ yields:
\begin{equation}
\| A \| \, \| B \| \, e^{- \frac{ \gamma s}{2}} \, \left[ 1 \, + \,
  \sqrt{ \frac{2}{ \pi \gamma s}} \, + \, 
 \frac{2 \, D_a(X,Y)}{ \pi \, \| \Phi \|_a \, C_a}  
e^{ \left(2 \, \| \Phi \|_a \, C_a \, + \frac{ \gamma}{2} \right) \, s} \, \right]
\end{equation}
As is demonstrated in (\ref{eq:dbd}), $D_a(X,Y)$ decays exponentially as
$e^{-a d(X,Y)}$. In this case, if we choose $s$ to be the solution of
the equation
\begin{equation} \label{eq:defs}
s \left( \, 2 \, \| \Phi \|_a \, C_a \, + \, \gamma / 2 \, \right) \, = \, a \, d(X, Y),
\end{equation}
then we have proven the result. Notice that we have chosen $\alpha$ 
in terms of $s$, which is defined independently of $b$, thus the 
condition $\gamma \geq 2 \alpha b$ will be satisfied for 
sufficiently small $b>0$.
\hfill \qed

In our proof of the Exponential Clustering Theorem above, we used several times 
the following useful fact, a proof of which appears in \cite{nachtergaele2005}.

\begin{lemma} \label{lem:ft} Let $E \in \mathbb{R}$, $\alpha >0$, and $z
  \in \mathbb{C}^+ = \{ z \in \mathbb{C} \, : \, \mbox{Im}[z]>0 \,
  \}$. One has that
\begin{equation}
\lim_{T \to \infty} \frac{1}{2 \pi i} \int_{-T}^T \frac{e^{iEt} e^{-
    \alpha t^2}}{t-z} dt \, = \, \frac{1}{2 \sqrt{ \pi \alpha}}
\int_0^{\infty} e^{i w z} e^{- \frac{(w-E)^2}{4 \alpha}} dw.
\end{equation}
Moreover, the convergence is uniform for $z \in \mathbb{C}^+$.
\end{lemma}

%%%%%%%%%%%%%%%%%%%%%%%%%%%%%%%%%%%%%%%%
%
%
%                                 The Lieb-Schultz-Mattis Theorem
%
%
%
%%%%%%%%%%%%%%%%%%%%%%%%%%%%%%%%%%%%%%%%%

\section{The Lieb-Schultz-Mattis Theorem} \label{sec:lsm}

As a final application of these locality bounds, in particular
both Theorem~\ref{thm:lr} and Theorem~\ref{thm:decay}, we were recently able to provide a
rigorous proof the Lieb-Schultz-Mattis theorem, see \cite{nasi}, which is valid in arbitrary
dimensions. In this section, we will discuss this result and outline the ideas 
which motivate our proof.

\subsection{The Result and Some Words on the Proof}
The classical Lieb-Schultz-Mattis Theorem (LSM), \cite{liscma}, concerns the
spin-1/2, anti-ferromagnetic Heisenberg chain. This model is defined
through a family of Hamiltonians $H_L$, acting on the Hilbert space
$\mathcal{H}_{[1,L]} = \bigotimes_{x \in [1,L]} \mathbb{C}^2$, with the form
\begin{equation} \label{eq:1dham}
H_L = \sum_{x=1}^{L-1} \vec{S}_x \cdot \vec{S}_{x+1}.
\end{equation}
Here, for each integer $x \in [1,L]$, the spin vectors $\vec{S}_x$
have components
\begin{equation}
S_x^j = \idty \otimes \cdots \otimes \idty \otimes S^j \otimes \idty \otimes
\cdots \otimes \idty, \quad
j =1,2,3,
\end{equation}
where $S^j$ is the corresponding spin-1/2 (Pauli) matrix acting on the
$x$-th factor of $\mathcal{H}_{[1,L]}$. The LSM Theorem may be stated as
follows.

\begin{theorem}(LSM, 1961) \label{thm:LSM} If the ground state of $H_L$ is unique, then
  the gap in energy between the ground state and the first excited
  state is bounded by $C/L$.
\end{theorem}

A further result by Lieb and Mattis in 1966, \cite{lima}, verified that
under certain conditions, for example when $L$ is even, the main
assumption in the LSM Theorem, specifically the uniqueness of the ground
state, is indeed satisfied. Almost twenty years later, the LSM Theorem
was generalized to encompass a variety of other one (and quasi-one)
dimensional models by Affleck and Lieb in \cite{afli}. In particular, this
result applies to those chains of even length with spins having 
arbitrary half-integer magnitude. Note: Here and in the rest of this 
section the term half-integer, or half-integral, refers to one-half of a positive 
odd integer, i.e., an element of the set $\mathbb{N} + 1/2$. 

For models to which the LSM Theorem applies, one expects that the
excitation spectrum corresponding to the thermodynamic limit has no
gap above the ground state energy. It is interesting to note that the
predictions of Haldane \cite{haldane83} suggest that such a result is 
rather sensitive to the type of interaction terms. In fact,  the spin-1,
anti-ferromagnetic Heisenberg chain is predicted to have a robust gap above the
ground state energy in the thermodynamic limit. 

In a work of 2004, see \cite{hastings2004}, Hastings argued that a higher dimensional
analogue of the LSM Theorem could be proven using the improved
locality bounds which have recently been established.  We will now summarize these
ideas and indicate how they may be implemented to demonstrate a rigorous proof of this
theorem. 

The multi-dimensional LSM Theorem, stated as Theorem~\ref{thm:lsmmd} below,
is valid for a large class of models; a detailed proof of this is contained in \cite{nasi}. 
For simplicity of presentation in this review article, we will restrict our attention 
to the spin-1/2, Heisenberg anti-ferromagnet, however, our general assumptions 
are discussed in Subsection~\ref{subsubsec:gen}. In $\nu$ dimensions, 
the model of interest is defined on subsets $V_L \subset \mathbb{Z}^{\nu}$ in analogy
to (\ref{eq:1dham}) above, i.e. one considers Hamiltonians
\begin{equation} \label{eq:nudham}
H_{L} = \sum_{\stackrel{x, y \in V_L:}{|x-y|=1}} \vec{S}_x \cdot \vec{S}_y
\end{equation}
acting on the Hilbert space $\mathcal{H}_{V_L} = \bigotimes_{x \in
  V_L} \mathbb{C}^2$. It is easy to state the new result.

\begin{theorem} \label{thm:lsmmd} If the ground state of $H_{L}$ is non-degenerate, then
  the gap, $\gamma_L$, above the ground state energy satisfies
\begin{equation}
\gamma_L \leq C \frac{\log(L)}{L}.
\end{equation}
\end{theorem}

The logarthmic correction which appears in Theorem~\ref{thm:lsmmd}, in contrast
to the original result of Theorem~\ref{thm:LSM}, seems to be an
inevitable consequence of the locality bounds we incorporate in our
proof. It is an interesting open question to determine whether or 
not there is a class of models, in dimensions $\nu >1$, for which one
can prove such a bound without the logarithmic correction. 

In essense, Theorem~\ref{thm:lsmmd} is proven using a variational argument. Letting $\psi_0$
denote the unique, normalized ground state, we know that for any
normalized vector $\psi_1$ with $|\langle \psi_0, \psi_1 \rangle| \neq
1$, the bound
\begin{equation}
0 < \gamma \leq \frac{\langle \psi_1, (H - E_0) \psi_1 \rangle }{1-|\langle \psi_0, \psi_1 \rangle|^2}
\end{equation}
is always valid. Here we have dropped the dependence of all quantities on the length
scale $L$. From this perspective, there are only three steps necessary
to prove the desired result. First, we must construct a normalized trial state $\psi_1$, 
as indicated above. Next, we must estimate the difference in the energy 
corresponding to $\psi_1$ and $E_0$. Lastly, we must ensure that the inner 
product $|\langle \psi_0, \psi_1 \rangle|$ remains sufficiently small.

This method of proof is complicated by the fact that the ground state
is virtually unknown, and therefore the means by which one should construct a trial state is
not apriori clear. Inspiration for the construction of our
variational state comes from the work of Hastings, again see \cite{hastings2004},
in which he proposes to consider the ground state of a modified
Hamiltonian, $H_{\theta}$, where the interactions in a given
hyperplane have been twisted by an angle of $\theta$; more on this below. 
The ground state
of this modified Hamiltonian may be regarded as the solution of a specific
differential equation, in the variable $\theta$, whose initial condition
corresponds to the unique ground state whose existence we assumed. The
solution of Hastings' differential equation is ameanable to analysis,
in particular, one can apply both the Lieb-Robinson bounds and the
Exponential Clustering Theorem to provide the desired energy and 
orthogonality estimates mentioned above. One may recall that the
clustering bounds, as in Theorem~\ref{thm:decay}, provide estimates which 
themselves depend on the size of gap $\gamma_L$. For this reason, the 
argument proceeds by way of contradiction. In fact, by assuming that 
there exists a sufficiently large constant $C$ for which the gap satisfies
$\gamma_L > C \log(L)/L$ for large enough $L$, we construct a trial state 
whose energy eventually violates this bound.

%%%%%%%%%%%%%%%%%%%%%%%%%%%%
%
%   More details.
%
%%%%%%%%%%%%%%%%%%%%%%%%%%%%%%%%%%

\subsection{A More Detailed Outline of the Proof} \label{subsec:more}

%%%%%%%%%%%%%%%%%%%%%%%%%%%%%%%%
%
%    The Trial State
%
%
%%%%%%%%%%%%%%%%%%%%%%%%%%%%%%%%%

\subsubsection{Constructing the Trial State} \label{subsubsec:ts}

In our proof, we use the fact that the Hamiltonians we
consider are assumed to have at least one direction of 
translation invariance. We incorporate this into our 
notation by considering finite subsets 
$V_L \subset \mathbb{Z}^{\nu}$ of the form 
$V_L = [1,L] \times V_L^{\perp}$ where we have isolated a particular
direction, which we will often refer to as the horizontal 
direction, and perpendicular sets $V_L^{\perp} \subset \mathbb{Z}^{\nu -1}$ 
with cardinality $| V_L^{\perp}| \leq C L^{\nu -1}$. For the orthogonality result,
we will also need to assume that $|V_L^{\perp}|$ is odd, see Subsection~\ref{subsubsec:est}.
The trial state is constructed from a perturbation of the Hamiltonian $H_{L}$ defined by
``twisting'' certain interaction terms. A twist in the hyperplane
situated at a site $m \in [1,L]$ is defined by replacing all interaction terms 
 $\vec{S}_x \cdot \vec{S}_y$ in (\ref{eq:nudham}) corresponding to horizontal bonds
 with $x=(m,v)$, $y=(m+1,v)$, and some $v \in V_L^\perp$ by terms of the form 
\begin{equation} \label{eq:twist}
h_{xy}(\theta)= \vec{S}_x\cdot e^{-i\theta S^3_y}\vec{S}_{y} e^{i\theta S^3_y}
\end{equation}
for some $\theta \in \mathbb{R}$. A doubly twisted Heisenberg
Hamiltonian is then given by 
\begin{equation} \label{eq:twistham}
H_{\theta,\theta^\prime}=\sum_{\stackrel{x, y \in V_L:}{| x- y |=1}}  h_{xy}(\theta_{xy})
\end{equation}
where 
\begin{equation}
\theta_{xy} = \begin{cases}\theta, & \text{if  $x=(m,v), y=(m+1,v)$ for some $v\in V_L^\perp$,}\\
\theta^\prime, & \text{if $x=(m+L/2,v), y=(m+1+L/2,v)$ for  $v\in V_L^\perp$,}\\
0, & \text{otherwise.} \end{cases}
\end{equation}
Here we have taken periodic boundary conditions in the horizontal direction. 

It is interesting to note the behavior of the lowest eigenvalues
of the singly twisted Heisenberg Hamiltonian $H_{\theta, 0}$ for a simple
spin ring with an even number of spins. The behavior depends in an interesting 
way on the magnitude of the spins. When the spins are half-integer, the two 
lowest eigenvalues cross at $\theta = \pi$. In contrast, when the spins are 
integer, they remain non-degenerate. The quasi-adiabatic evolution is a device
designed to construct a continuous path from the ground state of $H_{0,0}$ 
to the first excited state of $H_{2\pi,0}$ which, of course, are both identical 
to the unperturbed Hamiltonian.

\begin{figure}\label{fig:3eigs}
\begin{center}
\resizebox{!}{6truecm}{\includegraphics{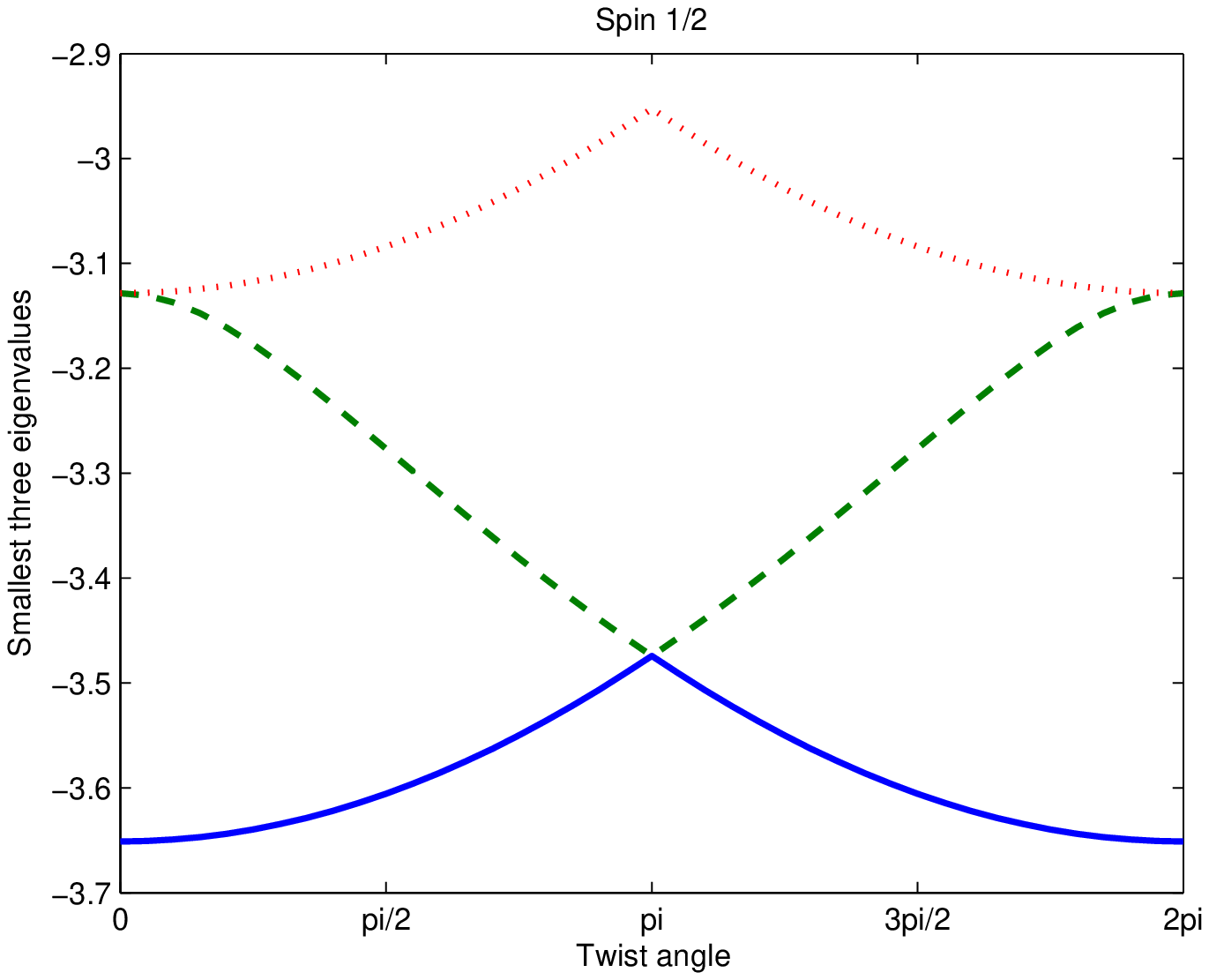}} 
\resizebox{!}{6truecm}{\includegraphics{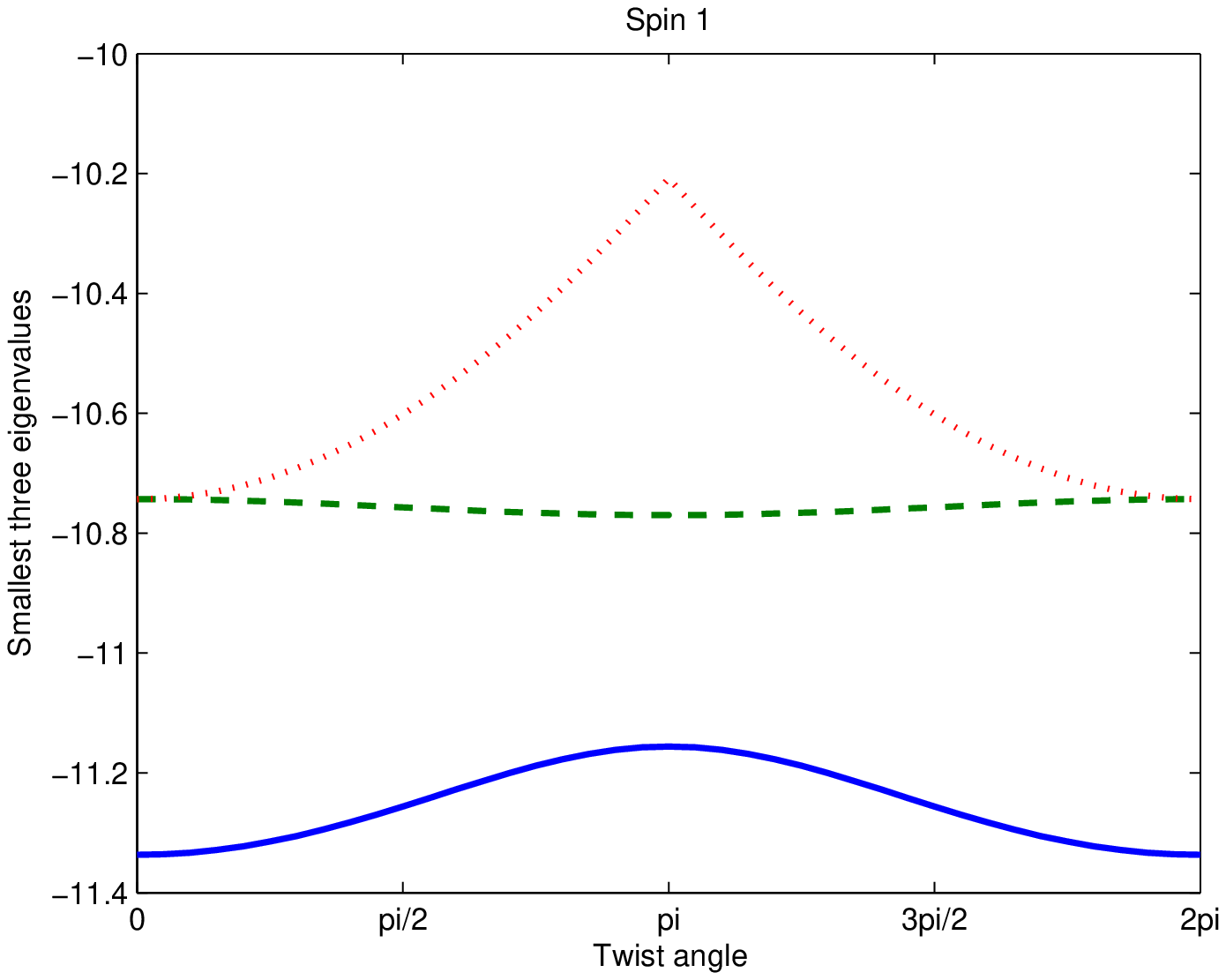}}
\end{center}
\caption{Plot of the three lowest eigenvalues of $H_{\theta, 0}$,
the singly twisted Heisenberg Hamiltonian for a ring of $8$ spins.
The magnitude of spins is $S=1/2$ in the plot on the left and $S=1$ 
on the right. }
\end{figure}

One of Hastings' crucial insights in \cite{hastings2004} is that, for the 
half-integer spin case, the first excited state can be obtained 
by applying a ``quasi-adiabatic evolution'' to the ground state; 
here $\theta$ is the evolution parameter. More concretely, let
$\psi_0( \theta, \theta')$ and $E_0( \theta, \theta')$ denote the
ground state and ground state energy of the doubly twisted Hamiltonian
$H_{\theta, \theta'}$, respectively. It is easy to see that along the path $\theta' = -
\theta$, the Hamiltonian $H_{\theta, -\theta}$ is unitarily equivalent
to the unperturbed Hamiltonian $H_L = H_{0,0}$. By differentiating
the eigenvalue equation $H_{\theta,-\theta}\psi_0(\theta,-\theta)
=E_0(\theta, - \theta) \psi_0(\theta,-\theta)$ and using that $\partial_\theta
E_0(\theta,-\theta) =0,$ one obtains 
\begin{equation} \label{eq:pertde}
\partial_{\theta} \psi_0(\theta, - \theta) = -\frac{1}{H_{\theta,-\theta} -
  E_0}[\partial_\theta H_{\theta,-\theta}]\psi_0(\theta, -\theta).
\end{equation}
Formally, equation (\ref{eq:pertde}) may be re-written using the 
Heisenberg dynamics as follows, 
\begin{equation} \label{eq:gsde}
\partial_{\theta} \psi_0(\theta, - \theta) = B( \theta) \psi_0(
\theta, - \theta),
\end{equation}
where the operator $B(\theta)$ is defined by
\begin{equation} \label{eq:bt}
B(\theta) =  -\int_0^\infty \tau_{it}(\partial_\theta
H_{\theta,-\theta})P_0(\theta, -\theta) \, dt.
\end{equation}
Here we have denoted by $\tau_t( \cdot) $ the dynamics generated by the Hamiltonian
$H_{\theta, -\theta}$, and $P_0( \theta, -\theta)$ is the corresponding
spectral projection onto the ground state. Due to the gap
assumption, (\ref{eq:bt}) is well-defined.

Equation (\ref{eq:gsde}) captures the evolution of the ground state
of a doubly twisted Hamiltonian along the path $\theta' = -\theta$
where the effect of the first twist is canceled by the second. 
The trial state is obtained by using an approximation of the
differential equation in (\ref{eq:gsde}) to describe the ground state
of the singly twisted Hamiltonian $H_{\theta, 0}$. 
Introduce the parameters $\alpha>0$ and $T>0$. To
approximate the imaginary time evolution corresponding to an
arbitrary Hamiltonian $H$, with dynamics $\tau_s( \cdot)$, at time $t>0$, we define
\begin{equation} \label{eq:ait}
A_{\alpha}(it,H)= \frac{1}{2\pi i}\int_{-\infty}^{+\infty} \tau_s(A)
\, \frac{e^{- \alpha s^2}}{s-it} \, ds
\end{equation}
for any observable $A$. In general, an anti-Hermitian operator of the form
\begin{equation} \label{eq:genbat}
B_{\alpha,T}(A,H)= -\int_0^T  [A_{\alpha}(it,H)-A_{\alpha}(it,H)^*] \, dt,
\end{equation}
will be used in place of (\ref{eq:bt}).

Note that the observable which is evolved in equation
(\ref{eq:bt}), $\partial_\theta H_{\theta,-\theta}$, contains 
terms of two types. The first are localized around the twist of angle
$\theta$ which correspond to horizontal bonds $(x,y)$ of the type 
$x=(m,v)$, $y=(m+1,v)$, and some $v \in V_L^{\perp}$. The second 
are similar, yet localized around the twist of angle
$- \theta$.  We group these two types of terms together and write
$\partial_{\theta}H_{\theta, -\theta} = \partial_1H_{\theta, -\theta}
- \partial_2 H_{\theta, -\theta}$ to simplify notation. By linearity, the operator
\begin{equation} \label{eq:linb}
B_{\alpha,T} \left( \partial_{\theta} H_{\theta, -\theta}, H_{\theta, -\theta} \right) \, = \,
B_{\alpha,T} \left( \partial_{1} H_{\theta, -\theta}, H_{\theta, -\theta} \right) \, - \, 
B_{\alpha,T} \left( \partial_{2} H_{\theta, -\theta}, H_{\theta, -\theta} \right).   
\end{equation}
Given a sufficiently large gap $\gamma_L$, there is a choice of the 
parameters $\alpha$ and $T$ for which the ground state 
$\psi_0( \theta, - \theta)$ is well approximated by the solution of the differential
equation (\ref{eq:gsde}) with $B(\theta)$ replaced by 
$B_{\alpha,T} \left( \partial_{\theta} H_{\theta, -\theta}, H_{\theta, -\theta} \right)$. 
Hastings' clever proposal is that one may also approximate the ground state of the singly 
twisted Hamiltonian, $\psi_0( \theta, 0)$, by evolving with just the first term on 
the right hand side of (\ref{eq:linb}) above. More concretely, consider the 
operator $B_{\alpha,T}(\theta)$ defined by setting 
$B_{\alpha,T}(\theta) = B_{\alpha,T}(\partial_1 H_{\theta, -\theta},
H_{\theta,-\theta})$ and solve the {\it Hastings' Equation} given by
\begin{equation} \label{eq:esde}
\partial_\theta\psi_{\alpha,T}(\theta)=B_{\alpha,T}(\theta) \psi_{\alpha,T}(\theta)
\end{equation}
with initial condition $\psi_{\alpha,T}(0)=\psi_0(0,0)$, i.e., the ground state of the unperturbed Hamiltonian.
Under the choice of parameters $\alpha=\gamma_L/L$ 
and $T=L/2$, the resulting variational state $\psi_1 =
\psi_{\alpha,T}(2\pi)$ may be estimated in such a way that Theorem~\ref{thm:lsmmd}
follows. We note that a particularly nice feature of the differential equation
(\ref{eq:esde}) is that the operator $B_{\alpha,T}(\theta)$ is
anti-Hermitian, and therefore the solution remains normalized for all
$\theta \in [0, 2 \pi]$. Hastings interprets the solution of (\ref{eq:esde}) as 
a {\it quasi-adiabatic evolution} of the ground state $\psi_0 = \psi_0(0,0)$.

%
%For convenience, we will use the notation 
%$\partial_{\theta}H_{\theta, -\theta} = \partial_1H_{\theta, -\theta}
%- \partial_2 H_{\theta, -\theta}$ to group these two types of terms together.
%Hastings' recipe for the trial state is now easy to state. 
%Consider the operator $B_{\alpha,T}(\theta)$ defined by setting 
%$B_{\alpha,T}(\theta) = B_{\alpha,T}(\partial_1 H_{\theta, -\theta},
%H_{\theta,-\theta})$ and solve the differential equation
%\begin{equation} \label{eq:esde}
%\partial_\theta\psi_{\alpha,T}(\theta)=B_{\alpha,T}(\theta) \psi_{\alpha,T}(\theta)
%\end{equation}
%with initial condition $\psi_{\alpha,T}(0)=\psi_0(0,0)$, i.e. choose
%the initial condition to be the ground state of the unperturbed Hamiltonian.
%Under the choice of parameters $\alpha=\gamma_L/L$ 
%and $T=L/2$, the resulting variational state $\psi_1 =
%\psi_{\alpha,T}(2\pi)$ may be estimated in such a way that Theorem~\ref{thm:lsmmd}
%follows. We note that a particularly nice feature of the differential equation
%(\ref{eq:esde}) is that the operator $B_{\alpha,T}(\theta)$ is
%anti-Hermitian, and therefore the solution remains normalized for all
%$\theta \in [0, 2 \pi]$. We refer to equation (\ref{eq:esde}) as {\it Hastings Equation}, 
%and he interprets it as a {\it quasi-adiabatic evolution} of the ground state $\psi_0 = \psi_0(0,0)$.

%%%%%%%%%%%%%%%%%%%%%%%%%%%%%%
%
%
%                         Locality of the Trial State
%
%
%%%%%%%%%%%%%%%%%%%%%%%%%%%%%%

\subsubsection{Locality and the Trial State} \label{subsubsec:loc}

One key technical lemma, which we use repeatedly in all of the
estimates that follow, concerns the local evolution of the 
solution to (\ref{eq:esde}). Consider a sub-volume $\Lambda_{L}(m) \subset V_L$
localized around the twist of angle $\theta$, for example, take
$\Lambda_{L}(m)$ to be of the form $[m-(L/4-2), m +(L/4-2)]
\times V_L^\perp$ which is slightly less than half of the system. 
Let $\rho_{\alpha, T}(\theta)$ and $\rho_0(\theta,-\theta)$ denote the density matrices corresponding
to the states $\psi_{\alpha,T}(\theta)$ and $\psi_0(\theta,-\theta)$.

\begin{lemma} \label{lem:locts}
Suppose there exists a constant $c>0$ such that $L\gamma_L\geq c$ and choose 
$\alpha = \gamma_L/L$ and $T=L/2$. Then, there exists constants $C>0$ and $k>0$ such
that
$$
\sup_{\theta\in[0,2\pi]} \left\|  {\rm Tr}_{V_L\setminus \Lambda_{L}(m)}\left[
\rho_{\alpha,T}(\theta) -\rho_0(\theta,-\theta)\right] \right\|_1
\leq C L^{2\nu} e^{-kL\gamma_L}.
$$
\end{lemma}

Since the proof proceeds by way of contradiction, the
assumption $L \gamma_L \geq c$ is part of the argument. 
Moreover, it is easy to produce a bound on $\gamma_L$ from
above that is independent of the length scale $L$. 
Lemma~\ref{lem:locts} demonstrates that if $L \gamma_L$ is 
sufficiently large, then the effect of ignoring the
second twist in the definition of $\psi_{\alpha, T}( \theta)$ is
negligible when one restricts their attention to observables localized
in $\Lambda_{L}(m)$. The proof of this lemma uses both the
Lieb-Robinson bound, Theorem~\ref{thm:lr}, and the Exponential Clustering
result, Theorem~\ref{thm:decay}.

%%%%%%%%%%%%%%%%%%%%%%%%%%%%%%%%%%%%%
%
%
%                       The estimates
%
%
%%%%%%%%%%%%%%%%%%%%%%%%%%%%%%%%%%%%%%%%

\subsubsection{The Estimates} \label{subsubsec:est}

{\it The Energy Estimate:}
To estimate the energy of the trial state, we consider the function
\begin{equation} 
E(\theta)= \langle \psi_{\alpha,T}(\theta), H_{\theta, -\theta}\psi_{\alpha,T}(\theta)\rangle.
\end{equation}
Due to the initial condition used to define $\psi_{\alpha, T}(
\theta)$, we know that $E(0)=E_0$ the ground state energy, and 
since $H_{2 \pi, - 2 \pi} = H_{L}$, $E(2\pi)$ corresponds to the 
energy of the trial state $\psi_1=\psi_{\alpha,T}(2\pi)$. The main 
idea here is to use the locality property of the trial state, i.e. 
Lemma~\ref{lem:locts}, and the unitary
equivalence of the Hamiltonians $H_{\theta, -\theta}$ to obtain an 
estimate on the derivative of this function which is uniform for
$\theta \in [0, 2\pi]$.
Explicitly, we can prove the following result.

\begin{theorem}
Suppose there exists a constant $c>0$ such that $L\gamma_L\geq c$ and choose 
$\alpha = \gamma_L/L$ and $T=L/2$. Then, there exists constants $C>0$ and $k>0$ such
that
$$
\vert  \langle \psi_1, H_{L}\psi_1\rangle -E_0\vert
\leq C L^{3\nu-1} e^{-kL\gamma_L}.
$$
\end{theorem}

{\it The Orthogonality Estimate:}
As we mentioned before, Hastings' quasi-adiabatic evolution
is norm preserving. In particular, we are guaranteed that $\Vert \psi_1\Vert
=\Vert \psi_0\Vert=1 $. Our argument that $\psi_1$ is sufficiently
orthogonal to $\psi_0$ makes essential use of the fact that the total
spin in each perpendicular set, $V_L^\perp$, is half-integer. In the 
case of the spin-1/2, anti-ferromagnetic Heisenberg model, this 
corresponds to the assumption that $|V_L^{\perp}|$ is odd. We have
the following theorem.

\begin{theorem} \label{thm:orth}
Suppose there exists a constant $c>0$ such that $L\gamma_L\geq c$ and choose 
$\alpha =\gamma_L/L$ and $T=L/2$. Then, there exists constants $C>0$ and $k>0$ such
that
$$
\vert  \langle \psi_1, \psi_0\rangle \vert
\leq C L^{2\nu} e^{-kL\gamma_L}.
$$
\end{theorem}

To prove this result we observe that, although the ground state
$\psi_0(\theta,-\theta)$ of the perturbed Hamiltonian is not
translation invariant, it is invariant with respect to ``twisted''  
translations. In fact, let $T$ be a unitary implementing 
the translation symmetry in the horizontal direction, specifically 
the direction in which we have imposed periodic boundary 
conditions, chosen such that $T\psi_0=\psi_0$. This is possible since 
$\psi_0$ is the unique ground state and the Hamiltonian is translation 
invariant with respect to $T$, i.e.,
$T^* H_{L} T = H_{L}$.

Define twisted translations by setting
\begin{equation} \label{eq:twisttrans}
T_{\theta,\theta^\prime}=T U_m(\theta) U_{m+L/2}(\theta^\prime)
\end{equation}
where the column rotation, $U_n(\theta)$, applies the rotation $e^{i\theta S^3_x}$ to all sites for the
form $x=(n,v)$, for some $v\in V_L^\perp$. The unitary equivalence of
the doubly twisted Hamiltonian $H_{\theta, -\theta}$ to $H_L =
H_{0,0}$ can also be expressed in terms of these column rotations
\begin{equation}
H_{\theta,-\theta} = W(\theta)^* H_{0,0} W(\theta), 
\end{equation}
where
\begin{equation}
W(\theta) =\bigotimes_{m<n\leq m+L/2} U_n(\theta).
\end{equation}
With these definitions, it is easy to see 
that $W(\theta)^* T W(\theta) = T_{\theta, -\theta}$ commutes with
$H_{\theta,-\theta}$, and therefore,
\begin{equation}
T_{\theta, -\theta}\psi_0(\theta,-\theta) =\psi_0(\theta, -\theta),
\end{equation}
as we claimed.

The main idea in the proof of Theorem~\ref{thm:orth} is to again use 
the locality properties of the solution of Hastings' Equation (\ref{eq:esde})
to show that 
\begin{equation}
T_{\theta,0}\psi_{\alpha,T}(\theta)\sim \psi_{\alpha,T}(\theta).
\end{equation}
Since the total spin in each $V^\perp_L$ is half-integer, again a
consequence of assuming $|V_L^{\perp}|$ is odd, the column
rotation $U_m(2\pi)=-\idty$. Clearly then, $T_{2\pi,0}=-T$, and
therefore, $T\psi_1 \sim -\psi_1$. As we have chosen $T$ so that, 
$T\psi_0 = \psi_0$, this implies that $\psi_1 $ is nearly
orthogonal to $\psi_0$.

%%%%%%%%%%%%%%%%%%%%%%%%%%%%%%%%%
%
%
%                        Generalizations
%
%
%
%%%%%%%%%%%%%%%%%%%%%%%%%%%%%%%%%%

\subsubsection{Generalizations} \label{subsubsec:gen}

As we previously indicated, the proof of the multi-dimensional 
Lieb-Schultz-Mattis Theorem, demonstrated in \cite{nasi}, applies 
to a large variety of models. In what follows below,
we will outline a list of assumptions which define a wide class
of Hamiltonians for which the LSM Theorem remains valid.

{\it The Basic Set-Up:} It is not important for our argument
that the underlying sets have a lattice structure, in particular the sets
$V_L$ need not be subsets of $\mathbb{Z}^{\nu}$. Rather, we
need only assume that there is, at least, one  
direction of increase, which previously we labeled the horizontal 
direction. We make this notion concrete by assuming that there exists 
an increasing sequence of finite sets $\{ V_L \}_{L \geq 1}$, exhausting 
some infinite set $V$, which are of the form $V_L =
[1,L] \times V_L^{\perp}$. We also assume a bound on the cardinality of the 
perpendicular sets of the form $|V_L^{\perp}| \leq c L^{\alpha}$ for
some $\alpha \geq 0$, and it is natural, but not necessary, to 
take $\alpha = \nu -1$. 

The interactions can also be of a general form. We assume that the set
$V$ is equipped with a metric $d$ and a function $F$ as described in 
Section~\ref{sec:lr}. To start with, we work with 
interactions $\Phi \in \mathcal{B}_a(V)$ for some $a>0$ so 
that the infinite volume dynamics is well defined from the beginning. In contrast
to (\ref{eq:nudham}), the more general finite volume Hamiltonians are of the
form
\begin{equation} \label{eq:genfvham}
H_{L} = \sum_{X \subset V_L} \Phi(X) + \mbox{ boundary terms}, 
\end{equation}
where we will assume periodic boundary conditions in the
horizontal direction and arbitrary boundary conditions in the other
directions. 

{\bf Assumption I.} Our first assumption is that the interaction $\Phi
\in \mathcal{B}_a(V)$ is translation invariant in the
horizontal direction. This is clearly the case for the Heisenberg
anti-ferromagnet, and in general, it means that for any $X
\subset V_L$,
\begin{equation}
\Phi(X + e_1) \, = \, \alpha_1 \left( \Phi(X) \right),
\end{equation}
where $X+e_1$ is the translation of all points in $X$ by one unit in
the horizontal direction and $\alpha_1(\cdot)$ is the translation automorphism which
maps $\mathcal{A}_{(n, V_L^{\perp})}$ into $\mathcal{A}_{(n+1,
  V_L^{\perp})}$ for all $n \in \mathbb{Z}$. Here the column sets 
  $(n, V_L^{\perp})$ are defined by $(n,V_L^{\perp}) = \{ x \in V_L : x =(n, v) \mbox{ for some } v \in V_L^{\perp} \}$. 
 Due to the assumed periodicity in the horizontal direction, this translation invariance can be
implemented by a unitary $T \in \mathcal{A}_{V_L}$, i.e. $\Phi(X+e_1)
= T^* \Phi(X) T$ for all $X \subset V_L$. This unitary $T$ will depend on
the length scale $L$, but we will suppress this in our notation.

{\bf Assumption II.} We further assume that the interaction has a
finite range $R>0$ in the horizontal direction. This assumption is 
not strictly necessary. It is clear from the estimates in \cite{nasi} that the
result remains true even if the interactions are of long range with
sufficiently fast decay.     

{\bf Assumption III.} We assume the interaction $\Phi$ has rotation
invariance about one axis. Again, this is clearly the case for the Heisenberg
model. In the more abstract setting, we specifically assume that for each $x \in V$ 
there is a local hermitian matrix, which we will denote by $S_x^3$, with
eigenvalues that are either all integer or all half-integer. These
matrices are also required to be translates of one another, i.e.,
for any $x \in V$, $\alpha_1(S_x^3) = S_{x+e_1}^3$. Rotation invariance for a general
interaction $\Phi$ means that for any $X \subset V_L$,
\begin{equation} \label{eq:rotinv}
U^*( \theta) \Phi(X) U( \theta) \, = \, \Phi(X) \, \mbox{ for all }
\theta \in \mathbb{R},
\end{equation} 
where the rotation $U( \theta)$ is defined by
\begin{equation} \label{eq:defrot}
U( \theta) \, = \, \bigotimes_{x \in V_L} e^{i \theta S_x^3}.
\end{equation}

{\bf Assumption IV.} We assume that the matrices introduced above,
i.e. $S_x^3$, are uniformly bounded in the sense that there exists a
positive real number $S$ for which $\sup_{x \in V} \| S_x^3 \| \leq
S$. In addition, we must assume an odd parity condition on the
spins. Define the parity $p_x$ of a site $x \in V$ to be 0 if the
eigenvalues of $S_x^3$ are all integers and 1/2 if they are all
half-integers. The odd parity assumption is that
\begin{equation}
\sum_{v \in V_L^{\perp}}p_{(n,v)} \in \mathbb{N} + 1/2,
\end{equation}
for all $n \in \mathbb{Z}$. For the spin-1/2 Heisenberg model, we
satisfied this assumption by taking the cardinality of the perpendicular 
sets, $|V_L^{\perp}|$, to be odd. In general, the sum of the spins over the perpendicular set
needs to be half-integer to ensure that the column rotations $U_n(\theta)$, as
defined after equation (\ref{eq:twisttrans}), satisfy $U_n( 2
\pi) = - \idty$. As we have seen, such an identity plays a crucial 
role in our argument for orthogonality.

{\bf Assumption V.} The ground state of $H_{L}$ is assumed to be
non-degenerate. In this case, it is also an eigenvector of the
translation $T$ and the rotations $U( \theta)$ introduced above. We
assume that the ground state has eigenvalue one for both $T$ and $U(\theta)$.

{\bf Assumption VI.} We assume that there are orthonormal bases of the
Hilbert spaces $\mathcal{H}_{V_L}$ with respect to which $S_x^3$ and
$\Phi(X)$ are real for all $x \in V$ and all finite $X \subset V$. This
assumption is satisfied by the Heisenberg model, and it is an
important symmetry which allows us, in general, to prove that the ground state
eigenvalue is invariant with respect to the doubly twisted Hamiltonian
$H_{\theta, -\theta}$.

The following theorem was proven in \cite{nasi}.

\begin{theorem} \label{thm:genlsmmd} Let $\Phi \in \mathcal{B}_a(V)$
  for some $a>0$. If $\Phi$ satisfies assumptions I-VI above, then 
  the gap, $\gamma_L$, above the ground state energy of $H_{L}$ satisfies
\begin{equation}
\gamma_L \leq C \frac{\log(L)}{L}.
\end{equation}
\end{theorem}

{\bf Acknowledgments:}  R.S. would like to acknowledge the gracious hospitality provided by
the Isaac Newton Institute for Mathematical Sciences where the writing of this work began. 
This article is based on work supported by the U.S. National Science
Foundation under Grant \# DMS-0605342. A portion of this work was also supported by the 
Austrian Science Fund (FWF) under Grant No.\ Y330.

\end{document}